\documentclass[twocolumn,showpacs,floatfix,prl]{revtex4}%
\usepackage{graphicx}%
\usepackage{amsmath}%
\usepackage{amsfonts}%
\usepackage{amssymb}
\usepackage{bm}

\def\d{{\partial}}

\def\e{{\epsilon}}
\def\0{{ {\bm 0} }}
\def\k{{ {\bm k} }}

\def\q{{ {\bm q} }}
\def\Q{{ {\bm Q} }}

\def\w{{\omega}}
\def\a{{\alpha}}
\def\b{{\beta}}

\def\g{{\gamma}}

\def\klim{{k\mbox{-}{\rm lim}}}
\def\wlim{{\w\mbox{-}{\rm lim}}}
\def\kwlim{{k(\w)\mbox{-}{\rm lim}}}
\allowdisplaybreaks[4]

\begin{document}
\title{
Linear Response Theory for Shear Modulus $C_{66}$
and Raman Quadrupole Susceptibility: \ 
Significant Evidence for Orbital Nematic Fluctuations in Fe-Based Superconductors
}
\author{
Hiroshi \textsc{Kontani} and 
Youichi \textsc{Yamakawa}
}

\date{\today }

\begin{abstract}
The emergence of the nematic order and fluctuations 
has been discussed as a central issue in Fe-based superconductors.
To clarify the origin of the nematicity, we focus on the 
shear modulus $C_{66}$ and the Raman quadrupole susceptibility 
$\chi_{x^2-y^2}^{\rm Raman}$.
Due to the Aslamazov-Larkin vertex correction,
the nematic-type orbital fluctuations are induced, and they enhances
both $1/C_{66}$ and $\chi_{x^2-y^2}^{\rm Raman}$ strongly.
However, $\chi_{x^2-y^2}^{\rm Raman}$ remains finite even at the 
structure transition temperature $T_S$, 
because of the absence of the band Jahn-Teller effect
and the Pauli (=intra-band) contribution,
as proved in terms of the linear response theory.
The present study clarifies that 
origin of the nematicity in Fe-based superconductors
is the nematic-orbital order/fluctuations.

\end{abstract}

\address{
Department of Physics, Nagoya University,
Furo-cho, Nagoya 464-8602, Japan. 
}
 

\sloppy

\maketitle

\begin{figure}[!htb]
\includegraphics[width=.85\linewidth]{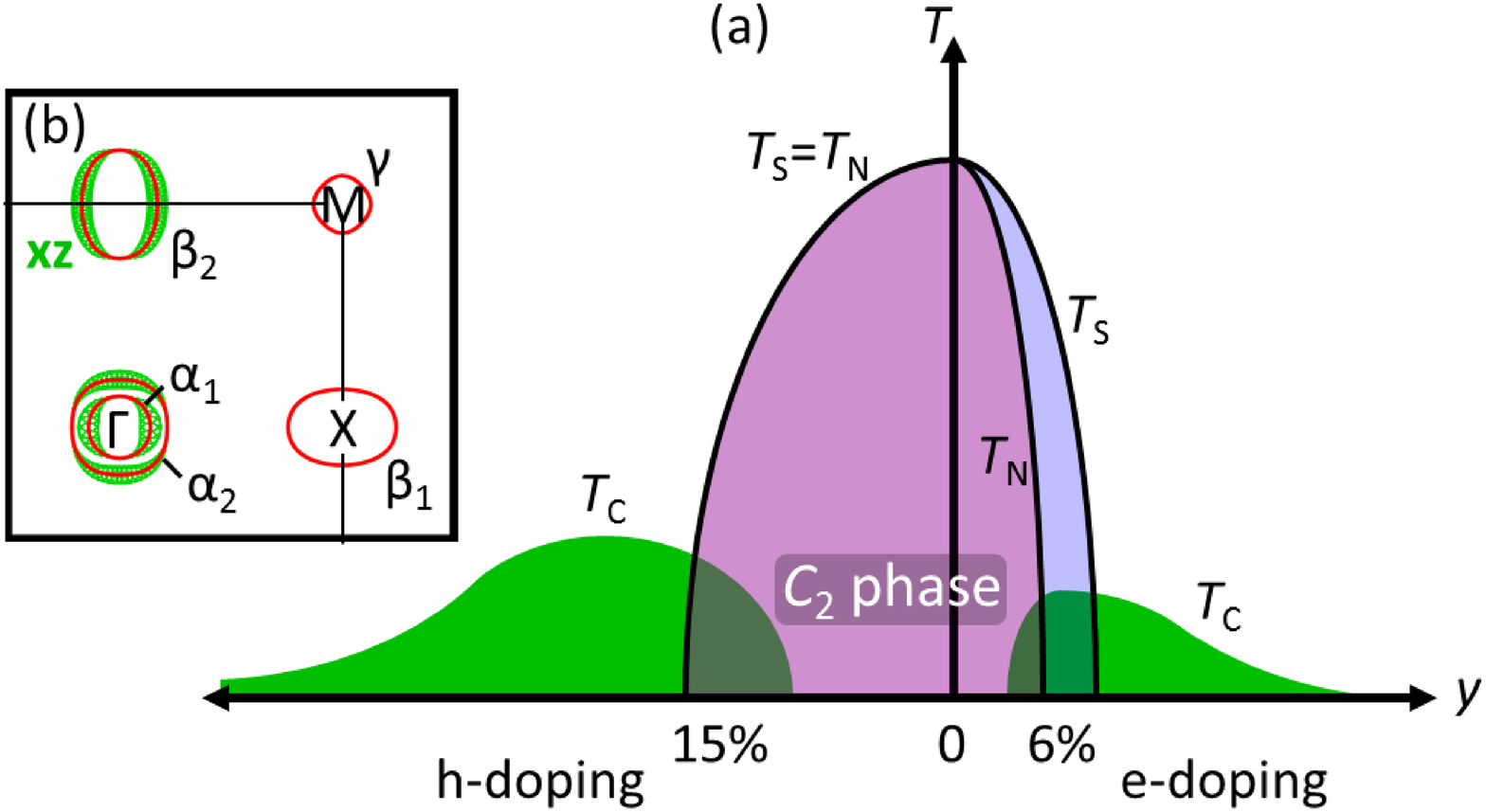}
\includegraphics[width=.95\linewidth]{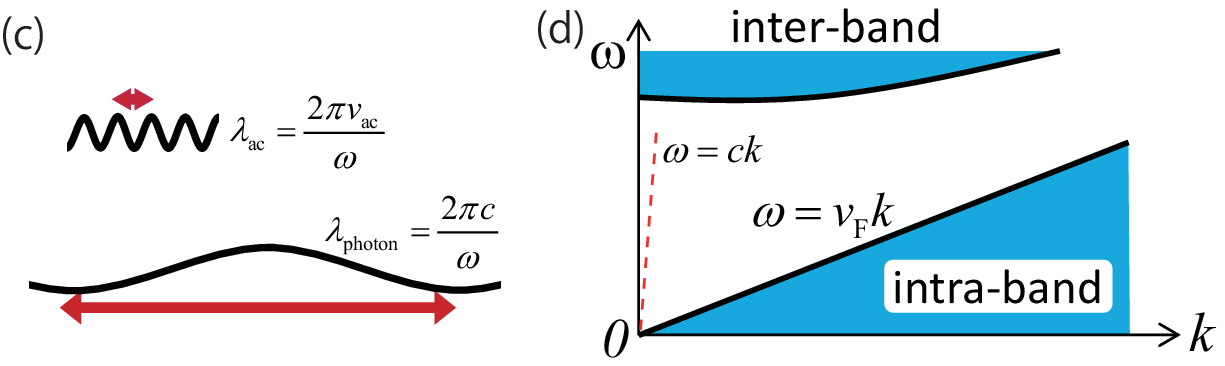}
\caption{(color online)
(a)Schematic phase diagram of Fe-based superconductors.
(b)Fermi surfaces for $y=0$.
The weight of $d_{xz}$ orbital is stressed by green circles.
(c) Relation $\lambda_{\rm photon}\gg\lambda_{\rm ac}$.
(d) Particle-hole excitation continuum.
}
\label{fig:FS}
\end{figure}

In Fe-based superconductors, 
the nematic order and fluctuations attract great attention
as one of the essential properties of the electronic states.
A schematic phase diagram of BaFe$_2$As$_2$ 
as a function of carrier doping $y$ is shown in Fig. \ref{fig:FS}:
For $y>0$ (e-doping), the non-magnetic orthorhombic ($C_2$) phase transition 
occurs at $T_S$, and the antiferro (AF) spin order is 
realized at $T_N (\lesssim T_S)$ in the $C_2$ phase.
In  Ba(Fe$_{1-x}$Co$_x$)$_2$As$_2$ ($y=x$),
both the structural and magnetic quantum critical points (QCPs)
are very close, and strong magnetic fluctuations are observed 
near the QCPs by NMR \cite{Imai}.
In addition, strong nematic susceptibility that couples to 
the $C_2$ structure deformation had been observed via the 
softening of shear modulus $C_{66}$
 \cite{Fernandes1,Yoshizawa,Yoshizawa-C33,Bohmer,Goto}
and in-plane anisotropy of resistivity \cite{Analytis}.
Similar softening of $C_{66}$ is also observed in 
(Ba$_{1-x}$K$_x$)Fe$_2$As$_2$ ($y=-x/2$; h-doping) \cite{Bohmer} 
and Fe(Se,Te) ($y=0$) \cite{Yoshizawa-11}.
Interestingly, in Ba(Fe$_{1-x}$Ni$_x$)$_2$As$_2$ ($y=2x$),
magnetic QCP and structural QCP are well separated,
and quantum criticalities are realized at both points
\cite{Zheng}.

Then, a natural question is what is the ``nematic order parameter''
that would be closely related to the pairing mechanism.
Up to now, both the spin-nematic mechanism \cite{Fernandes1} and 
ferro-orbital order mechanism \cite{Kruger,PP,WKu,Onari-SCVC}
had been proposed, and the softening of $C_{66}$
can be fitted by both mechanisms 
 \cite{Kontani-review,Fernandes2}.
The former predicts that the spin-nematic order 
$\langle {\bm s}_i\cdot{\bm s}_{i+{\hat x}} \rangle \ne0$ occurs above $T_N$
when the magnetic order $\langle {\bm s}_i\rangle$ is suppressed  
by the $J_1$-$J_2$ frustration.
As for the latter scenario, it was shown that the
orbital order $n_{xz}\ne n_{yz}$ is induced by spin fluctuations, 
due to strong spin-orbital mode-coupling given by the 
vertex correction (VC) \cite{Onari-SCVC,Ohno-SCVC,Tsuchiizu}.
The large $d$-orbital level splitting $E_{yz}-E_{xz}\sim 60$ meV
in the $C_2$ phase \cite{ARPES-Shen,Ding} 
may be too large to be produced by
spin nematic order via spin-lattice coupling.

Recently observed large quadrupole susceptibility $\chi_{x^2-y^2}^{\rm Raman}$
by electron Raman spectroscopy \cite{Gallais,Gallais2}
presents a direct evidence of the strong orbital fluctuations.
Although this result favors the orbital nematic scenario,
the observed enhancement of $\chi_{x^2-y^2}^{\rm Raman}$ 
is apparently smaller than the orbital susceptibility extracted from $C_{66}$.
For example, $\chi_{x^2-y^2}^{\rm Raman}$ remains finite at $T=T_S$, 
although $C_{66}^{-1}$ diverges at $T_S$.
Therefore, it should be verified 
whether both $C_{66}$ and $\chi_{x^2-y^2}^{\rm Raman}$
can be explained based on the orbital nematic scenario or not.

In this paper,
we analyze both $C_{66}$ and $\chi_{x^2-y^2}^{\rm Raman}$, both of 
which are key experiments to uncover the nematic order parameter.
It is found that both $C_{66}$ and $\chi_{x^2-y^2}^{\rm Raman}$ are enhanced by the 
orbital fluctuations due to Aslamazov-Larkin type VC (AL-VC).
However, $\chi_{x^2-y^2}^{\rm Raman}$ is less singular since
the band Jahn-Teller (band-JT) effect 
and the Pauli (=intra-band) quadrupole susceptibilities 
does not contribute to $\chi_{x^2-y^2}^{\rm Raman}$.
Since both $C_{66}$ and $\chi_{x^2-y^2}^{\rm Raman}$
are explained satisfactorily,
the orbital nematic scenario is essential for many Fe-based superconductors.

As for the pairing mechanism, at present, 
both the spin fluctuation mediated $s_\pm$ wave state
\cite{Kuroki,Hirschfeld,Chubukov}
and orbital fluctuation mediated $s_{++}$ wave state 
\cite{Kontani-RPA,Onari-SCVCS2} have been discussed.
When both fluctuations coexist,
nodal $s$-wave state can be realized \cite{Saito-Loop}.
The $s_{++}$-wave state is consistent with the robustness of $T_{c}$
against impurities
\cite{Onari-imp,Yamakawa-imp} and broad hump structure
in the inelastic neutron scattering 
\cite{Onari-neutron1,Onari-neutron2}.
The self-consistent vertex correction (SC-VC) method
\cite{Onari-SCVC, Onari-SCVCS2}
predicts the developments of ferro- and AF-orbital fluctuations,
and the freezing of the latter fluctuations would 
explain the nematic order at $T^*\sim200$K ($\gg T_S$)
\cite{Kasahara,Xray}.

First, we discuss the susceptibility at $\k\approx\0$
with respect to the quadrupole order parameter 
${\hat O}_{x^2-y^2} \equiv n_{xz}-n_{yz}$ in the Hubbard model.
For $U=U'+2J$, it is approximately given as 
\cite{Onari-SCVC,Ohno-SCVC}
\begin{eqnarray}
\chi_{x^2-y^2}(k)= 2\Phi(k)/(1-(U-5J) \Phi(k)) ,
\label{eqn:chiQ}
\end{eqnarray}
where $k=(\k,\w)$, and $\Phi(k) \equiv \chi^{(0)}(k)+X(k)$
is the intra-orbital (within $d_{xz}$ orbital) irreducible susceptibility:
$\chi^{(0)}(k)$ is the non-interacting susceptibility
and $X(k)$ is the VC for the charge channel.
The orbital nematic order  $n_{xz}\ne n_{yz}$ occurs 
when the charge Stoner factor $\a_c=(U-5J)\Phi(0)$ reaches unity,
which is realized near the magnetic QCP
since the AL-VC is proportional to the 
square of the magnetic correlation length
\cite{Onari-SCVC,Ohno-SCVC,Tsuchiizu}.

Next, we discuss the ``total'' quadrupole susceptibility in real systems,
by including the realistic quadrupole interaction due to the acoustic phonon 
for the orthorhombic distortion.
According to Ref. \cite{Kontani-soft}, it is given as
$-g_{\rm ac}(k) {\hat O}_{x^2-y^2}(\k){\hat O}_{x^2-y^2}(-\k)$, where 
${\hat O}_{x^2-y^2}(\k)$ is the quadrupole operator, and 
$g_{\rm ac}(k) = g\cdot (v_{\rm ac}|\k/\w|)^2/((v_{\rm ac}|\k/\w|)^2-1)$ 
is the phonon propagator multiplied by the coupling constants.
$v_{\rm ac}$ is the phonon velocity.
Since the Migdal's theorem tells that the effect of $g$ 
on the irreducible susceptibility is negligible,
the total susceptibility is
\begin{eqnarray}
\chi_{x^2-y^2}^{\rm tot}(k)= \chi_{x^2-y^2}(k)/(1-g_{\rm ac}(k)\chi_{x^2-y^2}(k)) .
\label{eqn:chi-tot}
\end{eqnarray}
%

Now, we discuss the acoustic and optical responses 
based on the total susceptibility (\ref{eqn:chi-tot}),
by taking notice that 
any susceptibilities in metals are {\it discontinuous} at $\w=|\k|=0$.
Since the elastic constant is measured under the static ($\w=0$) strain 
with long wavelength ($|\k|\rightarrow0$), $C_{66}$ is given as 
\begin{eqnarray}
C_{66}^{-1} \sim 1+\lim_{\k \rightarrow \0}g_{ac}(\k,0)\chi_{x^2-y^2}^{\rm tot}(\k,0)
= \frac{1}{1-g\chi_\klim} ,
\label{eqn:LR-C66} 
\end{eqnarray}
where $\chi_\klim \equiv \lim_{\k \rightarrow \0}\chi_{x^2-y^2}(\k,0)$ is called
the $k$-limit,
and the relation $g_{\rm ac}(k)=g$ for $\w=0$ is taken into account.
The structure transition occurs when $C_{66}^{-1}$ diverges.
When the AL-VC is negligible, $\chi_\klim$ is as small as $\chi_\klim^{(0)}$.
Even in this case, $C_{66}^{-1}$ can diverge when $g$ is very large,
which is known as the band-JT effect.
However, the band-JT mechanism cannot explain the 
strong enhancement of $\chi_{x^2-y^2}^{\rm Raman}$, as we will clarify later.
In fact, the fitting of experimental data in the present paper
indicates that the softening of $C_{66}$ is mainly given by the AL-VC:
The relation $1/g\sim \chi_\klim\gg\chi_\klim^{(0)}$ 
is satisfied in Fe-based superconductors.

Next, we derive the optical response in the DC limit,
measured by using the low-energy photon with
$k=(\k,\w=c|\k|)$ and $\w\rightarrow 0$.
Considering that the photon velocity $c$ is much faster than the 
Fermi velocity $v_{\rm F}$ and $v_{\rm ac}$, it is given as
\begin{eqnarray}
\chi_{x^2-y^2}^{\rm Raman} \sim \lim_{\w \rightarrow 0} \chi_{x^2-y^2}^{\rm tot}(\0,\w)
= \chi_\wlim ,
\label{eqn:LR-Raman}
\end{eqnarray}
where $\chi_\wlim \equiv \lim_{\w \rightarrow 0}\chi_{x^2-y^2}(\0,\w)$
is called the $\w$-limit \cite{Leggett,Kontani-M}.
Since $g_{\rm ac}(k)$ is zero for $|\w/\k|=c$, 
the band-JT effect does not contribute to the Raman susceptibility.
The physical explanation is that the acoustic phonons cannot be 
excited by photons because of the mismatch of the wavelengths
$\lambda_{\rm photon}\gg \lambda_{\rm ac}$ for the same $\w$
as shown in Fig. \ref{fig:FS} (c).
Also, since $c\gg v_{\rm F}$,
low-energy photon cannot induce the intraband particle-hole excitation
as understood from the location of the particle-hole continuum 
shown in Fig. \ref{fig:FS} (d).
This fact leads to the relationship 
``$\chi_\wlim$ is smaller than $\chi_\klim$''
as we discuss mathematically later.
For the charge quadrupole susceptibility,
this relationship holds even if the quasiparticle lifetime is finite
due to impurity scattering; see the Supplemental Material \cite{suppl}.
Therefore, $\chi_{x^2-y^2}^{\rm Raman}$ remains finite at $T\sim T_S$
although $C_{66}^{-1}$ diverges at $T_S$,
consistently with experiments \cite{Gallais,Gallais2}.

\begin{figure}[!htb]
\includegraphics[width=.99\linewidth]{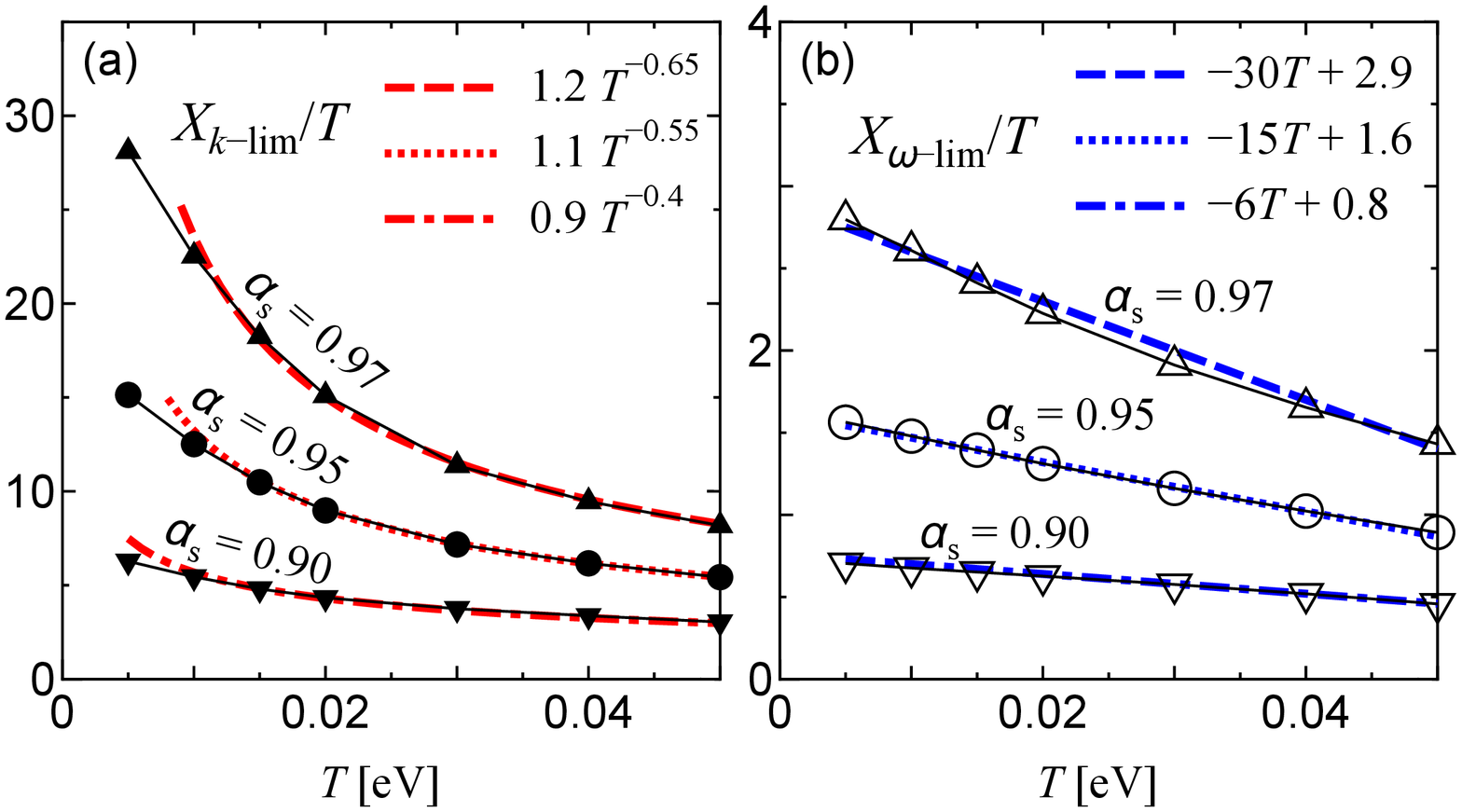}
\caption{(color online)
(a) $X_\klim/T$ and (b) $X_\wlim/T$ as functions of $T$.
Their $T$-dependences originates from $|\Lambda_Q^\kwlim|^2$
since $\xi^2\propto 1/(1-\a_s)$ is fixed.
}
\label{fig:numerical}
\end{figure}

Hereafter, we perform the numerical calculation of 
the quadrupole susceptibility in the five-orbital model.
The unit of energy is eV unless otherwise noted.
First, we discuss the $k$-limit and $\w$-limit of 
the bare bubble made of two $d_{xz}$-orbital Green functions.
They are connected by the following relation:
\begin{eqnarray}
\chi^{(0)}_\klim =\chi^{(0)}_\wlim+
\sum_\a^{\rm band} \left(-\frac{\d f_\k^\a}{\d\e_\k^\a}\right) \{z^{\a}_\k\}^2 ,
 \label{eqn:chi0k}
\end{eqnarray}
where $z_\k^{\a}=|\langle xz,\k|\a,\k\rangle|^2\ \le1$ 
is the weight of the $d_{xz}$-orbital on band $\a$, and 
$f_\k^\a = (\exp((\e_\k^\a-\mu)/T)+1)^{-1}$.
In Eq. (\ref{eqn:chi0k}),
$\displaystyle \chi^{(0)}_\wlim = \sum_{\a\ne\b}^{\rm band}
\frac{f_\k^\a-f_\k^\b}{\e_\k^\b-\e_\k^\a}z^{\a}_\k z^{\b}_\k$
is given by only the inter-band ($\a\ne\b$) contribution,
which is called the Van-Vleck term.
Therefore, $\chi^{(0)}_\klim$ is always larger than $\chi^{(0)}_\wlim$
due to the intra-band contribution
given by the second term in Eq. (\ref{eqn:chi0k}), called the Pauli term.
We obtain $\chi^{(0)}_\wlim\approx0.25$ and $\chi^{(0)}_\klim\approx0.45$
in the present model.

Next, we analyze AL-VC in detail,
since it is the main driving force of the orbital fluctuations.
The analytic expression of the AL term is given in 
Refs. \cite{Onari-SCVC,Ohno-SCVC}.
To simplify the discussion, we consider the intra-orbital 
(within $d_{xz}$-orbital) AL-term.
Then, $X_\kwlim$ is approximately given as
\begin{eqnarray}
X_\kwlim
&=& 3 T\sum_{\q}|\Lambda^\kwlim_\q|^2 V^s(\q,0)^2 ,
\label{eqn:Xkw}
\end{eqnarray}
where $\Lambda^\wlim_\q \equiv \lim_{\w\rightarrow0}\Lambda_q(\0,\w)$ 
and $\Lambda^\klim_\q \equiv \lim_{\k\rightarrow\0}\Lambda_q(\k,0)$ at $q=(\q,0)$:
$\Lambda_\q(\k,0)$ is the three-point vertex
made of three Green functions \cite{Onari-SCVC}.
Also, $V^s(q)=U+U^2\chi^s(q)$, where
$\chi^s(q)$ is the spin susceptibility for $d_{xz}$-orbital.
Here, we assume the following Millis-Monien-Pines form of $\chi^s(q)$
\cite{MMP}:
\begin{eqnarray}
\chi^s(q)=c \xi^2(1+\xi^2(\q-\Q)^2 +|\Omega_m|/\w_{\rm sf})^{-1} , 
\label{eqn:MMP}
\end{eqnarray}
where $\Q=(0,\pm\pi)$,
$\xi^2= l/(T-\theta)$ is the square of the spin correlation length,
and $\w_{\rm sf}=l'\xi^{-2}$ is the spin-fluctuation energy scale.
Quantitatively speaking, $X_\kwlim$ given by Eq. (\ref{eqn:Xkw}) is 
underestimated since non-zero Matsubara terms are dropped.
However, in the classical region $\w_{\rm sf}<2\pi T$,
which is realized in optimally-doped Ba(Fe,Co)$_2$As$_2$ \cite{CHLee},
$\chi^s(\q,\w_l)$ for $l\ne0$ is negligibly small.
In this case, we can safely use Eq. (\ref{eqn:Xkw}).

According to Eqs. (\ref{eqn:Xkw}) and (\ref{eqn:MMP}), we obtain
$X_\kwlim\sim T\{\Lambda^\kwlim_\Q\}^2\xi^2$
for two-dimensional systems.
$\Lambda^\klim_\q$ and $\Lambda^\wlim_\q$ are connected by the following relation:
\begin{eqnarray}
\Lambda^\klim_\q=\Lambda^\wlim_\q 
+\sum_{\a,\g}\sum_\k \left(-\frac{\d f_\k^\a}{\d\e_\k^\a}\right)
\frac{ \{z_\k^{\a} \}^2 z_{\k-\q}^{\g}}{\e_{\k-\q}^\g-\e_\k^\a}  ,
\label{eqn:L-k}
\end{eqnarray}
where $\Lambda^\wlim_\q$ is the inter-band Van-Vleck term \cite{Lambda-w}.
For $\q\approx\Q$, $\Lambda^\klim_\q$ increases strongly at low $T$,
because of the intra-band ``Pauli term`` 
in the second term of Eq. (\ref{eqn:L-k}).
Its main contribution is given by
$\a=\a_{1,2}$ and $\g=\b_2$ in Fig. \ref{fig:FS} (b).
Both Pauli and Van-Vleck terms are negative in the present model.
Therefore, the relationship $X_\klim > X_\wlim$ is satisfied.

Figure \ref{fig:numerical} (a) shows the temperature dependence of 
$X_\klim/T$ given by Eq. (\ref{eqn:Xkw}), by using the static 
RPA spin susceptibility $\chi^s(\q,0)$ 
obtained at $T=0.01$.
In this calculation, $\xi^2\propto (1-\a_s)^{-1}$ is fixed,
where $\a_s=U\chi^{(0)}(\Q,0)$ is the spin Stoner factor.
Thus, we obtain the relationship $X_\klim/T\sim T^{-0.5}\xi^2$, 
in which the factor $T^{-0.5}$ originates from the strong
$T$-dependence of $|\Lambda_\Q^\klim|^2$.
We also show the temperature dependence of $X_\wlim/T$ 
in Fig. \ref{fig:numerical} (b):
The relation $X_\wlim/T\sim (b-T)\xi^2$ is realized 
due to the $T$-dependence of $|\Lambda_\Q^\wlim|^2$ \cite{Lambda-w}.
Therefore, the relationship $X_\klim > X_\wlim$ 
is confirmed by the present calculation.

\begin{figure}[!htb]
\includegraphics[width=.99\linewidth]{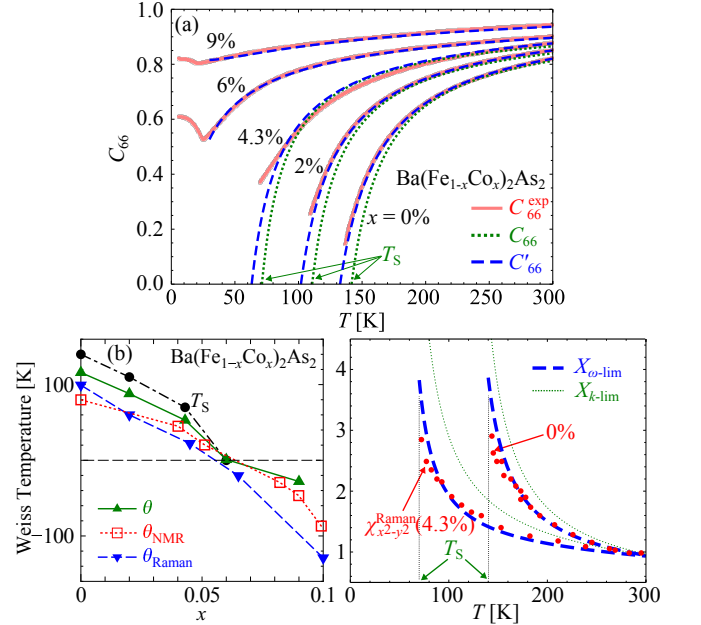}
\caption{(color online)
(a)  Fittings of the data $C_{66}^{\rm exp}$
normalized by the 33\% Co-Ba122 data in Ref. \cite{Bohmer},
shown by broad red lines.
The dotted lines $C_{66}$ is the fitting result
under the constraint $C_{66}=0$ at $T=T_S$,
and the broken lines $C_{66}'$ is the fitting without constraint.
(b) The Weiss temperature $\theta$ given by the present fitting.
$\theta_{\rm NMR}$ is the Weiss temperature of $1/T_1T$ \cite{Imai},
and $\theta_{\rm Raman}$ is given by the Raman spectroscopy \cite{Gallais}.
(c) $X_\klim$ and $X_\wlim$ given by the fitting of $C_{66}$.
Experimental data of $\chi_{\rm x^2-y^2}^{\rm Raman}$ 
are shown by red circles \cite{Gallais}. 
}
\label{fig:C66}
\end{figure}

Here, we perform the fitting of experimental data.
To reduce the number of fitting parameters,
we put $\chi_{x^2-y^2}\approx 2\Phi$ by assuming $(U-5J) \sim 0$,
which would be justified since the relation $J/U\sim0.15$ is predicted by the 
first principle study \cite{Miyake}.
Also, we put $\Phi\approx X$ by assuming that $X \gg \chi^{(0)}$.
Then, Eqs. (\ref{eqn:LR-C66}) and (\ref{eqn:LR-Raman}) are simplified as
\begin{eqnarray}
C_{66}^{-1} &\propto& 1/(1-2g X_\klim) ,
\label{eqn:LR-C66-2} \\
\chi_{x^2-y^2}^{\rm Raman} &\propto& X_\wlim ,
\label{eqn:LR-Raman-2}
\end{eqnarray}
where $X_\klim \equiv a_0 T^a \xi^2$ and $X_\wlim\equiv b_0(b-T)T\xi^2$:
According to Fig.\ref{fig:numerical}, $a\sim0.5$ and $b\sim0.1$ for $T>0.01$.

First, we fit the data of $C_{66}^{\rm exp}$,
which is normalized by the shear modulus due to phonon anharmonicity
(=33\% Co-Ba122 data) given in Ref. \cite{Bohmer}.
We putting $a=0.5$, and 
the remaining fitting parameters are $h=2g a_0l$ and $\theta$.
Figure \ref{fig:C66} (a)
shows the fitting result for Ba(Fe$_{1-x}$Co$_x$)$_2$As$_2$:
The ``dotted line $C_{66}$'' is the fitting result of 
$C_{66}^{\rm exp}$ under the constraint $C_{66}=0$ at $T=T_S$.
We fix $h=2.16$
for all $x$, and change $\theta$ from $116$K to $-30$K.
The ``broken line $C_{66}'$'' is the fitting for $x=0\sim0.09$
without the constraint, by using $h=2.67$.
Thus, both fitting methods can fit the $T$- and $x$-dependences of 
$C_{66}^{\rm exp}$ very well 
by choosing only $\theta(x)$ with a fixed $h$.
Figure \ref{fig:C66} (b) shows the obtained $\theta(x)$ 
by $C_{66}$-fitting ($x=0\sim0.043$)
and by $C_{66}'$-fitting ($x=0.06,0.09$), as explained above.
The obtained $\theta(x)$ is very close to 
$\theta_{\rm NMR}$ given by the Curie-Weiss fitting of $1/T_1T$ \cite{Imai},
which manifests the importance of the AL-VC.
Also, $\theta_{\rm Raman}$ is given by the Raman spectroscopy \cite{Gallais}.

\begin{figure}[!htb]
\includegraphics[width=.85\linewidth]{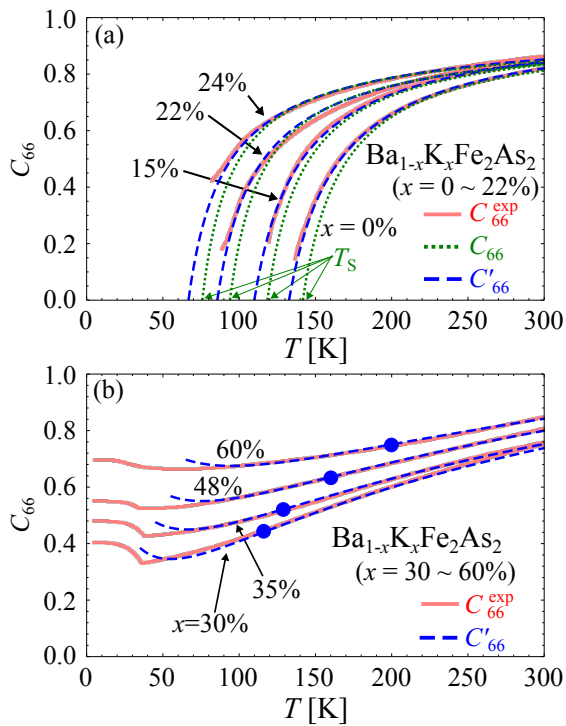}
\caption{(color online)
Fittings of shear modulus for (a) under-doped and (b) over-doped 
(Ba$_{1-x}$K$_x$)Fe$_2$As$_2$.
Experimental data $C_{66}^{\rm exp}$
are shown by broad red lines \cite{Bohmer}. 
}
\label{fig:C66-K}
\end{figure}

In Fig. \ref{fig:C66} (c),
we show $X_\klim$ obtained by the fitting of $C_{66}^{\rm exp}$
for Ba(Fe$_{1-x}$Co$_x$)$_2$As$_2$ at $x=0$ and $0.043$,
We also show $X_\wlim \sim X_\klim\cdot (b-T)T^{1-a}$
according to the numerical result in Fig. \ref{fig:numerical},
by putting $b=1400$K.
In  Fig. \ref{fig:C66} (c), all the data are normalized as unity at 300K.
Then, the relation $\chi_{x^2-y^2}^{\rm Raman} \sim X_\wlim$
is well satisfied, as expected from Eq. (\ref{eqn:LR-Raman-2}).
In addition, the relation $X_\wlim \ll X_\klim$ holds for $T\sim T_S$,
consistently with the report in Ba(Fe$_{1-x}$Co$_x$)$_2$As$_2$ 
\cite{Gallais}.

Figure \ref{fig:C66-K} (a) and (b)
shows the fitting results for (Ba$_{1-x}$K$_x$)Fe$_2$As$_2$ for
$x=0\sim0.24$ ($a=0.5$; $h=2.16$ 
for $C_{66}$ and $h=2.67$ 
for $C_{66}'$)
and $x=0.3\sim0.6$ ($a=0.58$; $h=4.98$ 
for $C_{66}'$), respectively.
In the present theory,
we can explain the existence of inflection points of $C_{66}$
in over-doped region (without structure transition)
reported experimentally \cite{Bohmer},
shown by large blue circles.
The inflection point originates from the factor $T^a$
in $X_\klim \propto T^{a}\xi^2$.
The fitting of over-doped data could be improved by
considering the deviation from the relation 
$X_\klim\propto T^{a}\xi^2$ at low $T$,
as recognized in Fig. \ref{fig:numerical} (a).
In addition, for $x\gtrsim0.5$, experimental pseudo-gap behavior of 
$1/T_1T \ (\propto \xi^2)$ below $\sim100$K \cite{Fukazawa}
would also be related to the inflection point of $C_{66}$.

In the present theory, we can fit $C_{66}^{\rm exp}$ very well for both
over-doped and under-doped regions in Ba$_{1-x}$K$_x$Fe$_2$As$_2$.
However, different set of parameters should be used in each region:
This fact indicates that the orthorhombic phase and superconducting phase
are separated by the first-order transition.
In fact, the $T^2$-like resistivity at the optimum doping $x\sim0.3$ 
indicates the absence of the orbital-nematic QCP in this compound.
We also note that the change in the topology of the electron-pockets,
called the Lifshitz transition, occurs in Ba$_{1-x}$K$_x$Fe$_2$As$_2$
near the optimal doping.

In this paper, we showed that Raman susceptibility at $\w=0$
is enlarged by the AL-VC.
The present theory predicts that the $\w$-dependence of the 
AC Raman susceptibility follows
$\chi_{x^2-y^2}^{\rm Raman}(\w)\sim X(\0,\w) \sim (1-i\w/\Gamma)^{-1}$, and 
$\Gamma$ is approximately $\sim\w_{\rm sf}$.
However, $\Gamma$ could be modified by the 
$\w$-dependence of $|\Lambda_q(k)|^2$.

In summary,
we presented a unified explanation for 
the softening of $C_{66}$ and enhancement of $\chi_{x^2-y^2}^{\rm Raman}$
based on the five-orbital model.
Both  $1/C_{66}$ and $\chi_{x^2-y^2}^{\rm Raman}$
are enhanced by the nematic-type orbital fluctuations induced by the AL-VC. 
However, $\chi_{x^2-y^2}^{\rm Raman}$ remains finite even at the 
structure transition temperature $T_S$, 
because of the absence of the band-JT effect
and the Pauli (=intra-band) contribution.
The present study clarified that the origin of the nematicity,
which is a central issue in Fe-based superconductors,
is the nematic-orbital order/fluctuations.

\acknowledgements
We are grateful to A.E. B\"{o}hmer for offering us her
experimental data published in Ref. \cite{Bohmer}.
We also thank Y. Gallais, A.V. Chubukov, J. Schmalian, R. Fernandes 
and S. Onari for useful discussions.
This study has been supported by Grants-in-Aid for Scientific 
Research from MEXT of Japan.
Part of numerical calculations were
performed on the Yukawa Institute Computer Facility.



\newpage

\section{[Supplemental Material]: 
Relationship $\chi_\klim>\chi_\wlim$ in the presence of impurities}

In the main text, we have studied the $k$-limit and $\w$-limit
of the quadrupole susceptibility $\chi_{x^2-y^2}(\q,\w)$, and 
found that the relationship $\chi_\klim > \chi_\wlim$ is satisfied.
The basis of this relationship is that 
the intra-band Pauli term is absent in both 
$\chi^{(0)}_\wlim$ and $X_\wlim \sim T\sum_\q |\Lambda^\wlim_\q|^2V^s(\q)^2$
in the absence of the elastic and inelastic scattering.
However, the relationship $\chi_\klim > \chi_\wlim$ is not trivial
when the scattering processes exist.
Here, we calculate both $\chi^{(0)}_\wlim$ and $\Lambda^\wlim_\q$
in the presence of the local nonmagnetic impurities
based on the $T$-matrix approximation in the five-orbital model.
For the charge quadrupole susceptivility, the relationship 
$\chi_\klim > \chi_\wlim$ is confirmed even in the presence of impurities.

We assume that the impurity potential $I$ is diagonal in the orbital basis.
(We write $d_{z^2}$, $d_{xz}$, $d_{yz}$, $d_{xy}$, $d_{x^2-y^2}$ orbitals
as $1,2, \cdots, 5$, respectively.)
Then, the $T$-matrix in the orbital basis is given as
\begin{eqnarray}
{\hat T}(\e_n)=I({\hat 1}-I\sum_\q {\hat G}(\q,\e_n))^{-1}
\end{eqnarray}
where $\e_n=(2n+1)\pi T$ and the Green function is 
${\hat G}(\q,\e_n)=(i\e_n+\mu-{\hat H}_\q^0-\Sigma^{\rm imp}(\e_n))$, and
\begin{eqnarray}
\Sigma^{\rm imp}(\e_n)\equiv n_{\rm imp}{\hat T}(\e_n)
\label{eqn:Sigma}
\end{eqnarray}
is the impurity self-energy when the impurity concentration is 
$n_{\rm imp} (\ll1)$.
The Bethe-Salpeter equation 
for the one-particle operator ${\hat O}$ is
\begin{eqnarray}
{\hat L}^{\rm imp}(k;\e_n)&=& {\hat O}
 +n_{\rm imp}\sum_\q {\hat T}(\w_l+\e_n){\hat G}(k+q)
\nonumber \\
& &\times {\hat L}^{\rm imp}(k;\e_n){\hat G}(q){\hat T}(\e_n)
\label{eqn:ThreeV}
\end{eqnarray}
where $q=(\q,\e_n)$ and $k=(\k,\w_l)$.
We will show the significant role of the VC given by the second term;
(${\hat L}^{\rm imp}-{\hat O}$).

First, we study the impurity effect on the bare-bubble $\chi^{(0)}(k)$
for the $O_{x^2-y^2}$ quadrupole.
The impurity effect is divided into the 
(i) self-energy correction (\ref{eqn:Sigma}) and 
(ii) vertex correction (\ref{eqn:ThreeV}).
If only (i) is taken into account, the bare-bubble
within the $d_{xz}$-orbital is given as
\begin{eqnarray}
\chi^{(0),\Sigma}(k)= -T \sum_q G_{2,2}(k+q)G_{2,2}(q)
\label{eqn:chiSig}
\end{eqnarray}
where $G$ includes the self-energy, and the suffix $2$ in $G$
represents the $d_{xz}$-orbital.
If both (i) and (ii) is taken into account, it is given as
\begin{eqnarray}
\chi^{(0),{\rm true}}(k)= -T \sum_{q,mm'} {\hat L}^{\rm imp}_{m,m'}(k;\e_n)
G_{m',2}(k+q)G_{2,m}(q)
\nonumber \\
\label{eqn:chiSigLam}
\end{eqnarray}
for ${\hat O}={\hat O}_{x^2-y^2}$ in Eq. (\ref{eqn:ThreeV}),
where $l,m=1\sim5$ represents the $d$-orbital.
$\chi^{(0),{\rm true}}$ gives the correct susceptibility for $n_{\rm imp}>0$,
whereas $\chi^{(0),\Sigma}$ is incorrect.

\begin{figure}[!htb]
\includegraphics[width=.9\linewidth]{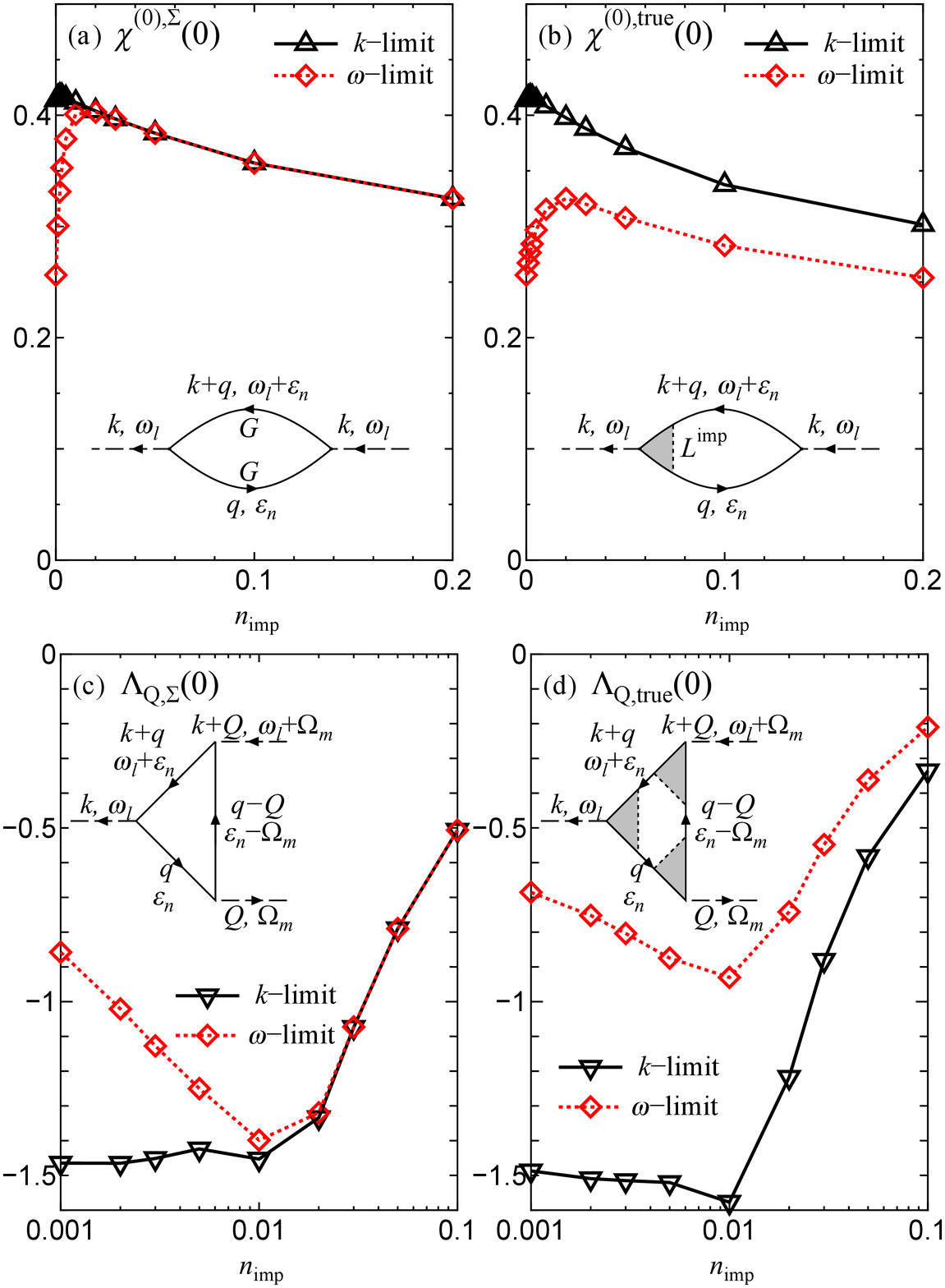}
\caption{(color online)
(a) $\chi^{(0),\Sigma}_\kwlim$, (b) $\chi^{(0),{\rm true}}_\kwlim$
(c) $\Lambda_{\Q,\Sigma}^\kwlim$, and (d) $\Lambda_{\Q,{\rm true}}^{\kwlim}$
as functions of $n_{\rm imp}$.
The correct results are given in (b) and (d).
Here, $\w$-limit values are obtained by extrapolating 
the data at $\w_l$ with $l=1\sim10$ to the real axis numerically.
}
\label{fig:AP}
\end{figure}

Here, we discuss the susceptibilities in the $k$-limit and $\w$-limit.
Using Eq. (\ref{eqn:chiSig}) or (\ref{eqn:chiSigLam}),
the former is simply given as $\chi_\klim=\chi(\k,\w_l)$ at $l=0$ and $\k=0$.
Here, we derive the latter numerically by extrapolating 
the data at $\w_l$ with $l=1\sim10$ to the real axis.
This procedure is successful at sufficiently low temperatures.
Figure \ref{fig:AP} (a) and (b) represent the numerically obtained
$\chi^{(0),\Sigma}_{\kwlim}$ and $\chi^{(0),{\rm true}}_{\kwlim}$ for $I=+1$, 
respectively.
We fix $T=3$ meV and $n=6.0$.
In (a), $\chi^{(0),\Sigma}_{\wlim}$ quickly increases with $n_{\rm imp}$,
and it is almost equal to $\chi^{(0),\Sigma}_\klim$ just for $n_{\rm imp}\gtrsim0.01$.
In (b), in contrast, $\chi^{(0),{\rm true}}_\wlim$ does not
reach the $k$-limit value even for  $n_{\rm imp}\sim0.1$.
In both (a) and (b), impurity effect on the $k$-limit value is very small.
Since $\chi^{(0),{\rm true}}_\kwlim$ gives the true susceptibility,
we conclude that the relationship 
$\chi^{(0)}_\klim>\chi^{(0)}_\wlim$
is satisfied even for $n_{\rm imp}>0$.

In Fig. \ref{fig:AP} (a), $\chi^{(0),\Sigma}_\wlim$ approaches to 
the $k$-limit value for $n_{\rm imp}>0$,
since the intra-band Pauli term also contributes to the $\w$-limit
($\k=0$ and $\w\rightarrow0$) 
due to the broadening of the quasiparticle spectrum caused by Im$\Sigma$.
However, the impurity three-point vertex
${\hat L}^{\rm imp}(k;\e_n)$ takes large value for 
$\w_l\cdot(\e_n+\w_l)<0$, and it suppresses the Pauli term.
These effects exactly cancel for conserved quantities:
For this reason, the charge and spin susceptibilities
become zero in the $\w$-limit even for $n_{\rm imp}>0$.
Although $O_{x^2-y^2}$ is not conserved,
the VC in ${\hat L}^{\rm imp}(k;\e_n)$ is nonzero in the present model,
and therefore the relationship 
$\chi^{(0),{\rm true}}_\klim>\chi^{(0),{\rm true}}_\wlim$ is satisfied.

Next, to discuss the AL-VC, we calculate 
$\Lambda_\Q^\kwlim$ at $\Q=(0,\pi)$ introduced in the main text
(Eq. (\ref{eqn:L-k}))
in the presence of impurities ($I=+1$), by which the AL-VC is given as 
$X_\kwlim\sim T |\Lambda_\Q^\kwlim|^2 \sum_\k \chi^s(\k)^2$.
Figure  \ref{fig:AP} (a) shows the numerically obtained
$\Lambda_{\Q,\Sigma}^{\kwlim}$, in which only $\Sigma^{\rm imp}$ is included.
We see that $\Lambda_{\Q,\Sigma}^{\wlim}$ increases with $n_{\rm imp}$,
and coincides with the $k$-limit value just for $n_{\rm imp}\gtrsim0.01$.
We also calculate $\Lambda_{\Q,{\rm true}}^{\kwlim}$, 
in which both $\Sigma^{\rm imp}$ and $L^{\rm imp}$ are taken into account properly.
In this case, $\Lambda_{\Q,{\rm true}}^{\wlim}$ does not reach the 
$k$-limit value even for $n_{\rm imp}\gtrsim0.1$
thanks to the VC in $L^{\rm imp}$.
Since $\Lambda_{\Q,{\rm true}}^{\kwlim}$ gives the true vertex function,
the relation of the AL-VC
$X_{{\rm true}}^{\klim}>X_{{\rm true}}^{\wlim}$
is confirmed even for $n_{\rm imp}>0$.

In summary, we confirmed that the relationship 
$\chi_\klim > \chi_\wlim$ is satisfied in the presence of impurities,
by taking both $\Sigma_{\rm imp}$ and $L^{\rm imp}$ into account correctly.
In other words, although the relation $\chi_\klim \approx \chi_\wlim$
is obtained by including $\Sigma_{\rm imp}$ only,
it is an artifact due to the neglect of the VC in $L^{\rm imp}$.
(Since $L^{\rm imp}={\hat O}$ for the charge current ${\hat O}={\bm v}_\k$,
such discontinuity will be absent for the conductivity.)
In real compounds, the Raman vertex ${\hat R}_{x^2-y^2}$
is very complex and momentum dependent.
In this paper, we take the momentum average of 
$R_{x^2-y^2}^{l,l}$ ($l=2,3$),
and consider the constant Raman vertex ${\hat O}_{x^2-y^2}$
to simplify the discussion.
In the present multiorbital model,
$L^{\rm imp}$ does not vanish even if the $\k$-dependence of
${\hat R}_{x^2-y^2}$ is taken into account,
so the relationship $\chi_\klim > \chi_\wlim$ should be satisfied 
for $n_{\rm imp}>0$.


\begin{thebibliography}{99}
\bibitem{Imai}
F. L. Ning, K. Ahilan, T. Imai, A. S. Sefat, M. A. McGuire, B. C. Sales, D. Mandrus, P. Cheng, B. Shen, and H.-H Wen, Phys. Rev. Lett. {\bf 104}, 037001 (2010)

\bibitem{Fernandes1}
R.M. Fernandes, L. H. VanBebber, S. Bhattacharya, P. Chandra, 
V. Keppens, D. Mandrus, M.A. McGuire, B.C. Sales, A.S. Sefat, 
and J. Schmalian, 
Phys. Rev. Lett. {\bf 105}, 157003 (2010). 

\bibitem{Yoshizawa}
M. Yoshizawa,  D. Kimura,  T. Chiba,  S. Simayi, Y. Nakanishi,  K. Kihou, 
C.-H. Lee,  A. Iyo,  H. Eisaki,  M. Nakajima, and  S. Uchida,
J. Phys. Soc. Jpn. {\bf 81}, 024604 (2012).

\bibitem{Yoshizawa-C33}
S. Simayi, K. Sakano, H. Takezawa, M. Nakamura, Y. Nakanishi, K. Kihou, M. Nakajima, C.-H. Lee, A. Iyo, H. Eisaki, S. Uchida, and M. Yoshizawa, J. Phys. Soc. Jpn. {\bf 82}, 114604 (2013).

\bibitem{Bohmer}
A. E. B\"{o}hmer, P. Burger, F. Hardy, T. Wolf, P. Schweiss, R. Fromknecht, M. Reinecker, W. Schranz, and C. Meingast, 
Phys. Rev. Lett. {\bf 112}, 047001 (2014).

\bibitem{Goto}
T. Goto, R. Kurihara, K. Araki, K. Mitsumoto, M.
Akatsu, Y. Nemoto, S. Tatematsu, and M. Sato, J. Phys.
Soc. Jpn. {\bf 80}, 073702 (2011).

\bibitem{Analytis}
H.-H. Kuo, J. G. Analytis, J.-H. Chu, R. M. Fernandes, J. Schmalian, and I. R. Fisher, Phys. Rev. B {\bf 86}, 134507 (2012).

\bibitem{Yoshizawa-11}
M. Yoshizawa, private communication.

\bibitem{Zheng}
R. Zhou, Z. Li, J. Yang, D. L. Sun, C. T. Lin, and G. Zheng, Nat. Commun {\bf 4}, 2265, (2013).

\bibitem{Kruger}
F. Kr\"{u}ger, S. Kumar, J. Zaanen, J. van den Brink, Phys. Rev. B {\bf 79}, 054504 (2009).

\bibitem{PP}
W. Lv, J. Wu and P. Phillips, Phys. Rev. B {\bf 80}, 224506 (2009);
W. Lv, F. Kruger, and P. Phillips,
Phys. Rev. B {\bf 82}, 045125 (2010).

\bibitem{WKu}
C.-C. Lee, W.-G. Yin, and W. Ku, 
Phys. Rev. Lett. {\bf 103}, 267001 (2009)

\bibitem{Onari-SCVC}
S. Onari and H. Kontani, Phys. Rev. Lett. {\bf 109}, 137001 (2012).  

\bibitem{Kontani-review}
H. Kontani, Y. Inoue, T. Saito, Y. Yamakawa and S.
Onari, Solid State Communications, {\bf 152}, 718 (2012).

\bibitem{Fernandes2}
R.M. Fernandes and  A.J. Millis, Phys. Rev. Lett. {\bf 111}, 127001 (2013).

\bibitem{Ohno-SCVC}
Y. Ohno, M. Tsuchiizu, S. Onari, and H. Kontani,
	J. Phys. Soc. Jpn. {\bf 82}, 013707 (2013).

\bibitem{Tsuchiizu}
M. Tsuchiizu, Y. Ohno, S. Onari and H. Kontani, 
Phys. Rev. Lett. {\bf 111}, 057003 (2013).

\bibitem{ARPES-Shen}
M. Yi, D. H. Lu, J.-H. Chu, J. G. Analytis, A. P. Sorini, A. F. Kemper, 
B. Moritz, S.-K. Mo, R. G. Moore, M. Hashimoto, W.-S. Lee, Z. Hussain, 
T. P. Devereaux, I. R. Fisher, and Z.-X. Shen,
Proc. Natl. Acad. Sci. USA {\bf 108}, 6878 (2011).

\bibitem{Ding}
H. Miao, L.-M. Wang, P. Richard, S.-F. Wu, J. Ma, T. Qian, L.-Y. Xing, X.-C. Wang, C.-Q. Jin, C.-P. Chou, Z. Wang, W. Ku, and H. Ding, arXiv.1310.4601

\bibitem{Gallais}
Y. Gallais, R. M. Fernandes, I. Paul, L. Chauviere, Y.-X. Yang, 
M.-A. Measson, M. Cazayous, A. Sacuto, D. Colson, and A. Forget,
Phys. Rev. Lett. {\bf 111}, 267001 (2013).

\bibitem{Gallais2}
Y.-X. Yang, Y. Gallais, R. M Fernandes, I. Paul, L. Chauviere, M.-A. Measson, M. Cazayous, A. Sacuto, D. Colson, and A. Forget 
 arXiv:1310.0934

\bibitem{Kuroki}
K. Kuroki, S. Onari, R. Arita, H. Usui, Y. Tanaka, H. Kontani, and H. Aoki,
Phys. Rev. Lett. {\bf 101}, 087004 (2008).

\bibitem{Hirschfeld}
P. J. Hirschfeld, M. M. Korshunov, I. I. Mazin 
Rep. Prog. Phys. {\bf 74}, 124508 (2011). 

\bibitem{Chubukov}
A. V. Chubukov, D. V. Efremov, and I. Eremin, Phys. Rev. B {\bf 78}, 134512 (2008).

\bibitem{Kontani-RPA}
H. Kontani and S. Onari, 
Phys. Rev. Lett. {\bf 104}, 157001 (2010).


\bibitem{Onari-SCVCS2}
S. Onari, Y. Yamakawa and H. Kontani, 
Phys. Rev. Lett. {\bf 112}, 187001 (2014).

\bibitem{Saito-Loop}
T. Saito, S. Onari and H. Kontani, Phys. Rev. B {\bf 88}, 045115 (2013).

\bibitem{Onari-imp}
S. Onari and H. Kontani, 
Phys. Rev. Lett. {\bf 103}, 177001 (2009).

\bibitem{Yamakawa-imp}
Y. Yamakawa, S. Onari, and H. Kontani, 
Phys. Rev. B {\bf 87}, 195121 (2013).

\bibitem{Onari-neutron1}
S. Onari, H. Kontani and M. Sato, 
 Phys. Rev. B {\bf 81}, 060504(R) (2010) 

\bibitem{Onari-neutron2}
S. Onari and H. Kontani,
Phys. Rev. B {\bf 84}, 144518 (2011) 

\bibitem{Kasahara}
S. Kasahara, H. J. Shi, K. Hashimoto, S. Tonegawa, Y. Mizukami, T. Shibauchi, K. Sugimoto, T. Fukuda, T. Terashima, A. H. Nevidomskyy, and Y. Matsuda, 
Nature {\bf 486}, 382 (2012).

\bibitem{Xray}
Y. K. Kim, W. S. Jung, G. R. Han, K.-Y. Choi, C.-C. Chen, T. P. Devereaux, A. Chainani, J. Miyawaki, Y. Takata, Y. Tanaka, M. Oura, S. Shin, A. P. Singh, H. G. Lee, J.-Y. Kim, and C. Kim, Phys. Rev. Lett. {\bf 111}, 217001 (2013).

\bibitem{Kontani-soft}
H. Kontani, T. Saito and S. Onari, 
Phys. Rev. B {\bf 84}, 024528 (2011).

\bibitem{Leggett}
P. Nozieres, {\it Theory of Interacting Fermi Systems} 
(Benjamin, New York, 1964);
A. A. Abrikosov, L. P. Gorkov and I. E. Dzyaloshinski, 
{\it Methods of Quantum Field Theory in Statistical Physics} 
Dover, New York, 1975);
A. J. Leggett, Phys. Rev. {\bf 140}, A1869 (1965).

\bibitem{Kontani-M}
H. Kontani, and K. Yamada, J. Phys. Soc. Jpn. {\bf 65}, 172 (1996);
H. Kontani, and K. Yamada, J. Phys. Soc. Jpn. {\bf 66}, 2232 (1997).

\bibitem{suppl}
H. Kontani and Y. Yamakawa, Supplemental Material

\bibitem{MMP}
A.-J. Millis, H. Monien and D. Pines, Phys. Rev. B {\bf 42}, 167 (1990); 
P. Monthoux and D. Pines,  Phys. Rev. B {\bf 47}, 6069 (1993).

\bibitem{CHLee}
P. Steffens, C.H. Lee, N. Qureshi, K. Kihou, A. Iyo, H. Eisaki, and M. Braden, 
Phys. Rev. Lett. {\bf 110}, 137001 (2013).

\bibitem{Lambda-w}
The analytic expression of $\Lambda^\wlim_\q$ is given as
%
\begin{eqnarray*}
\Lambda^\wlim_\q&=&
\sum_{\a,\b,\g}^{\a\ne\b} \sum_\k \left\{
\frac1{\e_\k^\b-\e_\k^\a}
\left(\frac{f_\k^\b}{\e_\k^\b-\e_{\k-\q}^\g}-\frac{f_\k^\a}{\e_\k^\a-\e_{\k-\q}^\g} 
\right) \right.
 \nonumber \\
&+&\left. \frac{f_{\k-\q}^\g}{(\e_{\k-\q}^\g-\e_\k^\a)(\e_{\k-\q}^\g-\e_\k^\b)}
 \right\} z_\k^{\a} z_\k^{\b} z_{\k-\q}^{\g}
 \nonumber \\
&+&\sum_{\a,\g}\sum_\k \frac{f_{\k-\q}^\g - f_\k^\a}{(\e_\k^\a-\e_{\k-\q}^\g)^2}
\{z_\k^{\a}\}^2 z_{\k-\q}^{\g} .
\end{eqnarray*}
%

\bibitem{Miyake}
T. Miyake, K. Nakamura, R. Arita and M. Imada, 
J. Phys. Soc. Jpn. {\bf 79}, 044705 (2010).

\bibitem{Fukazawa}
M. Hirano, Y. Yamada, T. Saito, R. Nagashima, T. Konishi, T. Toriyama, 
Y. Ohta, H. Fukazawa, Y. Kohori, Y. Furukawa, K. Kihou, C.-H. Lee, 
A. Iyo, and H. Eisaki, J. Phys. Soc. Jpn., {\bf 81}, 054704 (2012).


\end{thebibliography}
\end{document}